\documentclass[letterpaper,useAMS,usenatbib,usegraphicx]{mn2e}

\usepackage{amssymb}

\def\h{^{\rm h}}
\def\m{^{\rm m}}
\def\s{^{\rm s}}

\def\HI{\rm H\,{\sc i}}
\def\OVII{O\,{\sc vii}}
\def\OVIII{O\,{\sc viii}}
\def\mjb{mJy beam$^{-1}$}
\def\jb{Jy beam$^{-1}$}
\def\k{km s$^{-1}$}
\def\pp{^{\prime\prime}}
\def\cm2{cm$^{-2}$}
\def\c3{cm$^{-3}$}

\title[\HI \ observations of RX J0822--4300 in Puppis A]
{Observations of the neutral hydrogen surrounding the radio quiet 
neutron star RX J0822--4300 in Puppis A}
\author[E. M. Reynoso et al.]{E. M. Reynoso,$^{1,2}$\thanks{E-mail:
ereynoso@physics.usyd.edu.au}\thanks{Member of the Carrera del 
Investigador Cient\'\i fico, CONICET, Argentina.}\thanks{Postdoctoral 
External Fellow of CONICET, Argentina.} A. J. Green,$^{1}$ S. 
Johnston,$^{1}$ G. M. Dubner,$^{2}$\footnotemark[2] E. B. 
Giacani,$^{2}$\footnotemark[2]\newauthor and W. M. Goss,$^{3}$ \\
$^{1}$School of Physics, University of Sydney, NSW 2006, Australia\\
$^{2}$Instituto de Astronom\'\i a y F\'\i sica del Espacio, CC 67,
Suc 28, 1428 Buenos Aires, Argentina\\
$^{3}$National Radio Astronomy Observatory, P. O. Box 0,
Socorro, New Mexico 87801, USA\\}
\begin{document}

\date{Accepted . Received ; in original form }

\pagerange{\pageref{firstpage}--\pageref{lastpage}} \pubyear{2003}

\maketitle

\label{firstpage}

\begin{abstract}

We have observed the \HI \ distribution in an area $40^\prime \times 
40^\prime$ around the neutron star candidate RX J0822--4300, which is 
located in the supernova remnant Puppis A. The observations of the 
$\lambda21$ cm line were obtained with the Australia Telescope Compact 
Array (ATCA) and were combined with single dish data from the Southern 
Galactic Plane Survey. The spatial resolution is $90\pp$, and the velocity 
resolution, 1 \k . A sensitivity of $\sim 0.7$ K was achieved.
The results revealed a double lobed feature of reduced emission at
+16 \k, centered on the central compact object (CCO), and aligned with 
an \HI \ hole blueshifted by 13 \k .  The \HI \ depressions have probably 
been created by the sweeping up of $\sim 2$ M$_\odot$. The alignement
between the lobes and the optical expansion centre of Puppis A suggests
that the CCO could be ejecting two opposite jets. The velocity at which 
the two lobes are best defined allowed us to confirm that the distance to 
Puppis A is 2.2 kpc, based on a systemic velocity of +16 \k . The hydrogen 
column density computed using this systemic velocity is consistent with 
estimates from models for X-ray spectra, thus reinforcing our conclusion 
that the kinematic distance is 2.2 kpc. 

\end{abstract}


\section{Introduction}

In recent years, different X-ray missions have revealed a new class of
unresolved objects with no radio counterpart and very high X-ray to optical
flux ratios. Most of these objects are found  in the interior of supernova 
remnants (SNR). \citet{b47} propose four categories for these exotic sources:
sources showing pulsations with periods between 6 and 12 seconds are
called ``anomalous X-ray pulsars'' (AXP), or ``soft gamma-ray repeaters''
(SGR) if bursts of $\gamma$-ray emission have been detected. The remaining
sources are called ``central compact objects'' (CCO) if they are located in
the interior of an SNR, and ``isolated neutron stars'' (INS) if they are not
associated with a known SNR. 

The nature of CCOs is still not clear. They have been interpreted as young, 
energetic radio pulsars with unfavourable beaming \citep{b3}, as neutron stars 
(NS) with long initial periods and high magnetic fields \citep{b34,b12}, or as 
fast spinning objects with low magnetic fields \citep{b20}. Their X-ray 
emission is generally explained as thermal radiation from cooling NSs (e.g 
Zavlin, Tr\"umper \& Pavlov 1999), with typical temperatures of a few 10$^6$ 
K, inferred from their thermal spectral characteristics. A review 
of CCOs has recently been published by \citet{b47}.
 
An \HI \ study of the interstellar medium (ISM) towards the SNR G296.5+10.0 
revealed that its associated CCO, 1E 1207.4--5209, lies near the centre of a 
small, elongated depression, about $5\farcm 3$ in diameter, or 3.2 pc at an 
assumed distance of 2.1 kpc \citep{b21}. The depression has a small line width, 
less than 2 \k , and is deepest at $v\simeq -16$ \k , in agreement with the 
systemic velocity of the SNR. \citet{b21} suggest that the X-ray flux from 1E 
1207.4--5209 heats the local gas, providing a hot background against which the 
colder, foreground \HI \ is seen in absorption. Motivated by this striking 
discovery, we began a search for similar traces in the neutral gas around 
other candidate NSs, both isolated or associated with SNRs. In this paper, 
we report the results obtained for RX J0822-4300, the CCO located in the 
interior of Puppis A.
 
Puppis A (G260.5--3.4) is the remnant of a Type-II supernova explosion which
ocurred approximately 3,700 years ago \citep{b39}. At radio wavelengths,
this remnant appears as a distorted shell (Figure 1; Dubner et al. 1991),
flattened to the east presumably due to the interaction with an external
cloud \citep{b8,b30}. The most widely accepted distance of 2.2 kpc is 
derived from the possible association with this cloud at +16 \k . However, 
a recent study based on the OH 1667 MHz line (Woermann, Gaylard \& Otrupcek 
2000) suggests that the distance to Puppis A could be somewhat smaller, at
about 1.3 kpc.
 
\begin{figure}
\includegraphics[width=252pt]{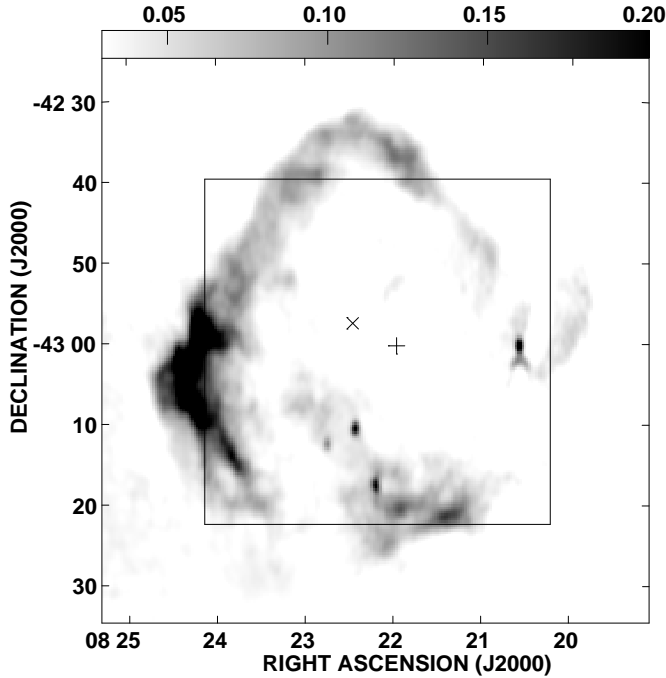}
 \caption{Radio continuum mosaic of Puppis A at 1415 MHz obtained with 
the Very Large Array \protect\citep{b9}. The angular resolution is $77\pp \times
43\pp$. The grayscale is in \jb. The inner frame shows the area covered 
by the present survey. The plus sign indicates the position of RX 
J0822--4300, while the cross shows the optical expansion centre computed 
by \protect\citet{b39}.}
\end{figure}

In X-rays, Puppis A appears as an open shell \citep{b28} with a
flattening to the east perfectly interlocking with the eastern cloud at
+16 \k \ (Figure 7 in Reynoso et al. 1995). ROSAT and ASCA observations
have revealed an X-ray source near the centre of the shell (Petre, Becker 
\& Winkler 1996). This compact object, RX J0822--4300, is located $6\farcm 
1$ to the southwest of the kinematical centre of expansion as determined 
from the proper motion of the fast optical filaments \citep{b38,b39}.
The lack of optical or radio counterparts down to the detection limits 
strongly suggests that this X-ray source is a NS, while its location 
suggests a physical association with the SNR Puppis A.
 
Pavlov, Zavlin \& Tr\"umper (1999) reported a marginal detection of pulsations
with a periodicity of $P\simeq 75$ ms. However, this result has been retracted
by \citet{b47}. A deep search did not detect an associated radio synchrotron 
nebula (Gaensler, Bock \& Stappers 2000a), which may indicate that RX 
J0822--4300 belongs to a population distinct from young radio pulsars. X-ray 
spectra from RX J0822--4300 are consistent with those predicted for NSs
with a hydrogen atmosphere, having an effective temperature, $T_{eff}=(1.6-1.9) 
\times 10^6$ K and \HI \ column density, $N_{\rm H}=(2.9-4.7)\times 10^{21}$ 
\cm2 \ \citep{b42}. This temperature is consistent with standard NS cooling 
models.
 
\begin{figure}
\vskip 8 cm
\caption{Average \HI \ spectra based on ATCA data (a) toward the surveyed 
area, and (b) toward RX J0822--4300. Gaussian fits to the highest peak near 
$v=+13$ \k \ (excluding the dip in Fig. 2b; see text) are shown as dashed 
lines. The sum of all fitted Gaussians appears as a dash-dotted line. The 
residuals are shown by dotted lines.}
\end{figure}

In this paper, we present new \HI \ observations carried out with the
Australia Telescope Compact Array (ATCA) of a field of about $40^\prime
\times 40^\prime$ (inner square in Fig. 1) around the compact source RX
J0822--4300. Our goal is to study the \HI~ distribution and kinematics
around the CCO, looking for signs of interaction between this source 
and the surrounding ISM. This study also allows us to revisit 
the problematic distance to Puppis A.

\section{Observations and data reduction}

The interferometric observations were obtained with the ATCA during two
sessions of 7.5 and 2 h with the 750D array (recording visibilities
from baselines 76.5 to 735 m), on 2001 April 27 and 28 respectively, 
and one session of 12 h with the EW 367 array (with baselines 46 to 
367 m) on 2002 March 23. The pointing centre was RA$=08\h 22\m 10\s$, 
Dec.$= -43^\circ 01^\prime 0\farcs 0$ (J2000), $\sim 2\farcm 5$ away 
from the position of RX J0822--4300 (RA$=08\h 21\m 57\fs 5$, Dec.$= 
-43^\circ 00^\prime 15\farcs 8$; J2000). The correlator configuration 
used has 1024 channels for a total bandwith of 4 MHz, centered at 1420 
MHz. The corresponding velocity resolution is 1 \k .  A continuum image 
was obtained simultaneously with a bandwidth of 128 MHz centered at 1384 
MHz. The source PKS B1934--638 was used for flux density and bandpass 
calibration. To calibrate phases two sources were used: PKS B0748--440 
for the 2001 sessions, and PKS B0823--500 for the 2002 session.
 
The data were processed with the MIRIAD software package \citep{b32}.
For the first two hours of the 2001 April 27 session, shadowing affected
the correlated data between antennas 1 and 2, and they were deleted. 
To subtract the continuum component from the \HI \ data set, a linear 
baseline was fitted to 700 line-free channels. The final \HI \ cube was 
constructed using the MIRIAD routine INVERT with a 20$\pp$ cell size, 
retaining 250 channels from --30 to +176 \k. Sidelobes were suppressed over 
an area of $42\farcm 5 \times 42\farcm 5$. To optimise detection of \HI 
\ in emission, only spatial frequencies shorter than 1.5 k$\lambda$ (315
m) were used. The \HI \ data were convolved with a $\sim 90\pp$ beam and
cleaned to a level of 5$\sigma$ (60 \mjb), where $\sigma$ is the noise of 
the line free channels in the dirty image.
The continuum image was constructed with the same geometry and angular 
resolution as the \HI \ data and a sensitivity of $\sigma \simeq 6$ \mjb 
\ (0.5 K) was achieved in the cleaned image.

\begin{figure}
\vskip 4 cm
\caption{Grayscale and contour image of the average \HI \ emission within 
5.4 \k \ of $v=+16.1$ \k . The brightness temperature scale is shown at the
top of the image, in units of K. The contours vary from 63.7 to 97.3 K in
steps of 2.1 K (3$\sigma$). The plus sign indicates the position of RX 
J0822--4300. The cross shows the position calculated by \protect\citet{b39} for 
the optical expansion centre. The beam, $90\pp \times 90\pp$, is plotted 
as a white open circle in the bottom right corner. The noise level is 
$\sim 0.7$ K.}
\end{figure}

To recover structures with the shortest spatial frequencies, the ATCA \HI 
\ data were combined in the {\it u,v} plane with single dish (Parkes
telescope) data from the Southern Galactic Plane Survey (SGPS; 
McClure-Griffiths et al. 2001), using the IMMERGE routine. No tapering was 
applied to the low resolution cube. The rms of the combined cube is $\sim 
0.7$ K ($\sim 10$ \mjb) per channel. 

\section {Results}

To isolate \HI \ absorption due to RX J0822--4300, we
compared two profiles: one averaged over the whole area surveyed, and the 
other taken in a direction towards the CCO. These profiles are shown in
Figure 2. In Fig. 2a, the highest emission peak (at $v<+20$ \k ) can 
be reproduced by the sum of two Gaussian components: one near +3.5 \k , 
probably corresponding to the local Orion arm, and a broader, stronger 
one, peaking at $\sim +14$ \k. The local distribution of \HI \ at this 
Galactic longitude is subject to confusion due to the Gum nebula 
\citep{b29,b41}.
 
Fig. 2b looks remarkably similar to Fig. 2a except for a dip in the \HI \
emission near $v=+16$ \k. If the Gaussian distributions shown in Fig. 2a 
are reproduced in Fig. 2b, the residuals show a negative spike centered 
at +16.1 \k \ with a spectral width of 5.4 \k . The central velocity of
the dip is coincident with the systemic velocity deduced
for Puppis A based on a VLA \HI \ study \citep{b30}. At this velocity,
an \HI \ cloud was found to be interacting with the eastern flank of 
the radio continuum shell. This cloud was also detected in CO \citep{b8}.

\begin{figure*}
\vskip 14 cm
\caption{Grayscale images of the \HI \ emission towards Puppis A within
the velocity range +7.9 to +20.3 \k , after combining interferometric ATCA
observations with single dish data from the SGPS \protect\citep{b24}. The brightness 
temperature scale is shown on the top of the first row of channels, in 
units of K. The area covered is the same as in Figs. 3 and 5. The beam, $90\pp 
\times 90\pp$, is plotted in the bottom left corner of the first channel. 
Velocities are shown at the top right corner of each channel. The position 
of RX J0822--4300 is indicated by a cross. For clarity, the average \HI \ 
emission has been subtracted from each of the channel maps.}
\end{figure*}

Figure 3 displays the average \HI \ distribution within 5.4 \k \ of
$v=+16.1$ \k . The image shows a pronounced density gradient, weaker
in the west. The distribution is consistent with the existence of the 
eastern cloud and with the morphology of the SNR radio shell. In addition 
to the \HI \ cavity opening to the west, there is an elongated minimum
$\sim 15^\prime$ long, oriented in the NE-SW direction, with a position
angle of about 60$^\circ$. Strikingly, RX J0822--4300 (indicated by a plus 
sign) lies near the centre of this \HI \ feature. Two local minima appear
inside the elongated feature, one at each side of the CCO. The western 
minimum is more extended and pronounced than the eastern one.  This 
linear depression has a length of about 8.5 pc at the assumed distance 
to the SNR of 2.2 kpc, with an average width of 1.6 pc. 

\citet{b39} found that the expansion centre of the O-rich fast filaments 
in Puppis A lies at RA$=08\h 20\m 44\fs 3$, Dec.$= -42^\circ 47^\prime 
48\pp$ (B1950). In Fig. 3, this position is indicated by a cross. If the 
expansion centre of the optical filaments pinpoints the site of the SN 
explosion, it is significant that the eastern lobe is aligned with the 
path traced from this location to the CCO.

The combination of this evidence suggests that this \ double-lobed \HI \ 
structure is not only related to Puppis A but, more specifically, to its CCO. 
It is well known that \HI \ emission is widespread and it is usual to find 
structures on all scales at different radial velocities within the Galaxy. For 
that reason, we have made a careful search of the whole \HI \ cube for
features morphologically similar to the double lobed structure.  To 
facilitate this search, an averaged profile over the whole area was computed 
and subtracted from the \HI \ data. As an example, in Figure 4 we show a set 
of channels from 7.9 to 20.3 \k ,  after subtraction of the average profile. 
We did not find any feature similar to the linear depression at $\sim +16$ \k. 
Hence, it is likely that this elongated structure is related to Puppis A which
has a systemic velocity of +16 \k . Moreover, the symmetry of the lobes around 
RX J0822--4300 suggests that the CCO is physically connected to this structure. 
This can be explained if the gas inside the remnant has a neutral component.
There are some mechanisms by which this could happen. \citet{b49} show that 
if a SNR is interacting with a dense external medium, as is the case of 
Puppis A, the shock can propagate through the intercloud medium leaving 
behind the clouds, still cold and at rest. These clouds gradually evaporate, 
and in middle-aged SNRs (like Puppis A), the hot post-shock gas can co-exist 
with dust at less than 100 K. The neutral hydrogen phase of the ISM is
typically found at cool temperatures, in the range $\sim 50 - 150$ K, or 
at warm temperatures of several thousand K. For SNR G296.5+10.0, internal 
neutral gas is the most probable explanation for the \HI \ depression in 
which the associated CCO is immersed \citep{b21}. Assuming that the reduced 
emission from the lobes is due to absence of material, we estimate that 1.7 
M$_\odot$ of neutral hydrogen must have been evacuated.  

The inspection of the whole data cube revealed another small depression 
between +2.2 and +3.8 \k , coincident with RX J0822--4300. Figure 5
shows an average of the emission within this velocity interval. The CCO 
appears to be slightly off-centre from the \HI \ hole (hereafter
``central hole'') by $37\pp$ (0.4 pc at a distance of 2.2 kpc), 
analogous to the results for 1E 1207.4--5209 \citep{b21},
which is off-centre by almost 1 pc. Most of the other minima present in
the field are produced through absorption of background continuum emission.
This is not the case for the central hole, where no continuum emission is
detected to a limit of 20 \mjb \ or 1.5 K. It is reasonable to propose that 
RX J0822--4300 has a direct influence on this \HI \ depression. The alignment 
observed between this minimum and the lobes (Figure 4) is remarkable. 
However, it is not clear whether the coincidence in position between the 
central hole and RX J0822--4300 is a projection effect.

\begin{figure}
\vskip 4 cm
\caption{Gray-scale and contour image of the average \HI \ emission 
between +2.2 and +3.8 \k . The brightness-temperature scale is shown 
at the top of the image, in units of K. The contours vary from 57.4 to 
82.6 K in steps of 2.1 K (3$\sigma$). The plus sign indicates the position 
of RX J0822--4300. The beam, $90\pp \times 90\pp$, is plotted as a white 
open circle in the bottom right corner. The noise level is $\sim
0.7$ K.}
\end{figure}

\begin{figure}
\includegraphics[width=242pt]{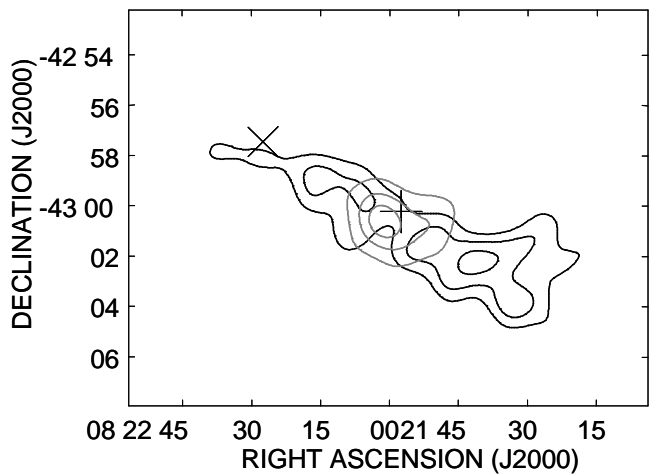}
\caption{Sketch of the \HI \ contours that depict the two lobes in Fig. 3 
(thin, black lines). The plus sign indicates the position of RX J0822--4300, 
while the cross stands for the optical expansion centre computed by 
\protect\citet{b39}. 
The thick, grey lines delineate the central hole in Fig. 5.}
\end{figure}

If we assume that the distance to the central hole is the same as the
distance to the lobes, 2.2 kpc, the mean diameter of the hole is 1.2
pc ($1\farcm 9$). The ratio between the minor and major axis is 0.7. If
we assume that the minimum emission at the hole is due to an absence
of neutral hydrogen, then the depleted mass amounts 0.3 M$_\odot$. 
 
\section {Discussion}

The present \HI \ observations have revealed an intriguing structure
consisting of two aligned depressions like lobes. The western one is
broader and deeper than the eastern lobe, consistent with effects
expected from the general density gradient of gas at the distance of
Puppis A \citep{b8,b30}. 
In the previous section,  we pointed out that the elongated \HI \ 
feature has : (a) the same systemic velocity as that of Puppis A, (b) 
alignment with the compact source RX J0822--4300, which lies between 
the two lobes, and (c) the same orientation as a line from the optical 
expansion centre of Puppis A to the CCO. These three points, together 
with the absence of similar features at other velocities, strongly 
suggest that this \HI \ structure is associated with RX J0822--4300, 
either related to the past history (translation from the expansion 
centre to the present location) or with currently ongoing activity 
(e.g. mass ejection). However, the central hole, although positionally
coincident with RX J0822--4300 (Fig. 5), is at a different velocity
to the lobes.

\subsection {The origin of the {\HI} structure}

\citet{b3} propose that radio quiet NSs are normal pulsars which are
not beamed 
towards the observer. From a theoretical point of view, the rotational 
energy loss of a young pulsar is converted into a relativistic wind of 
electron-positron pairs coupled with the ambient magnetic field which, 
when confined by the high-pressure inside a SNR, generates a synchrotron 
emitting nebula. At present, 13 out of 15 pulsars associated with Galactic
SNRs are known to have created pulsar wind nebulae (PWN) \citep{b18,b46,b48},
while H$\alpha$ bow-shock nebulae have been detected around five NSs 
(Chatterjee \& Cordes 2002 and references therein). It is expected that 
the local distribution of neutral gas might show the effect of the injection 
of particles and energy from the neutron star. This could be manifest as
ionization of gas or bulk sweeping up of low density gas to the boundaries
of the possible jets or outflows. The absence of detected radio continuum
emission from ionized gas supports to the second hypothesis.
 
\begin{figure}
\vskip 4 cm
 \caption{X-rays ROSAT image of Puppis A \protect\citep{b27} in gray-scale,
superimposed onto the \HI \ contours of Fig. 3. The RQNS is shown enclosed 
by an open triangle.}
\end{figure}

We have compared the \HI \ structure with the ROSAT image of Puppis A
(Figure 7). No X-ray enhancement can be detected at the position of the
\HI \ depressions, as would be expected if the gas had been heated to
$\gtrsim 10^6$ K. Moreover, the X-ray emission appears to be enhanced
along the periphery of the \HI \ structure, mainly at the end of the
eastern lobe.  

The morphology strongly suggests that the lobes represent
jets or collimated outflows ejected by RX J0822--4300. Synchrotron
nebular tails interpreted as jets associated with pulsars have been 
detected both in X-rays and in radio continuum (e.g. PSR B1757--24, 
Kaspi et al. 2001). In some cases, like PSR B1509--58 in G320.4--1.2 
\citep{b16,b19}, the radio continuum is enhanced at both sides of the 
X-ray features. In these cases, the radio synchrotron emission is 
thought to originate in a cylindrical sheath around a hot, thermal 
X-ray jet \citep{b43,b13}, where particle acceleration might occur 
via laterally expanding shocks. Such lateral shocks could account 
for the \HI \ emission feature detected at an anomalous velocity 
\citep{b44} aligned with the X-ray jet in G320.4--1.2. In most cases, 
the collimated outflows are believed to be directed along the pulsar's 
spin axis \citep{b22,b19}. The process by which jets are produced is 
not well understood.

\citet{b10} discovered two aligned lobes in radio continuum centered 
on the Vela pulsar, with a morphology very similar to the double
lobed \HI \ structure detected here. Two opposing X-ray jets are 
aligned in the direction of the proper motion of the Vela pulsar 
\citep{b51,b52,b50}.  However, the radio continuum double lobed structure 
is almost prependicular to this axis \citep{b10}. The fact that the 
jets align with the proper motion is not coincidental, and has 
implications for formation models of neutron stars. The radio 
luminosity of the two lobes is only $10^{-6}$ the spin down energy 
$\dot E$ of the pulsar. The morphological similarity between the 
double lobed structures in Vela and RX J0822--4300 is suggestive 
of a common origin, yet their different orientation with respect 
to the proper motion of the NS is difficult to interpret.

There are also problems in trying to interpret the two \HI\ lobes as 
jets. First, jets have velocities of at least a few thousand \k \ 
\citep{b13,b23}, which seems incompatible with the narrow spectral 
width of the \HI \ structure. Also, jets are usually less than 1 pc 
in extension, unlike the \HI \ lobes. Rather, the lobes are comparable 
in size with typical PWN \citep{b2}. It has been suggested that the 
motion of the pulsar might determine the morphology of the surrounding 
ISM with the pulsar lying at the apex of a nebula emitting in radio
or X-rays (e.g. W44, Frail et al. 1996; IC 443, Olbert et al. 2001). 
Such a scenario for RX J0822--4300 is unlikely since it cannot
explain either the western lobe or the central hole. However, it is 
possible that two effects are combined: the ejection of two opposing 
jets into a medium in turn modified by the passage of the CCO.  

Undoubtedly, the main difficulty with trying to explain the double 
lobed feature as an energetic pulsar phenomenon is the lack of any 
significant radio or X-ray counterpart. If RX J0822--4300 were a 
normal pulsar, the non-detection of a radio continuum counterpart 
to the \HI \ structure could be explained by the inefficiency in 
converting $\dot E$ into detectable radio emission. The lack of 
detection of any pulsation or PWN associated with RX J0822--4300, 
probably (but not necessarily) means that this source is not a high
$\dot E$ pulsar. Developing a detailed explanation to account for 
the \HI \ double lobed structure discovered here constitutes a challenge 
that could help to interpret the nature of unconventional neutron stars.

\subsection {The distance to Puppis A}
 
The distance to Puppis A of $2.2 \pm 0.3$ kpc was determined based on a 
systemic velocity of +16 \k \ obtained in a VLA \HI \ absorption study 
\citep{b30}.  Additional determinations based on the proper motion of 
optical filaments \citep{b6} and on a possible association with Vela OB1 
\citep{b31}, yielded a distance of 1.9 kpc, in good agreement with the
previous value. However, an OH 1667 MHz study \citep{b40} casts 
doubt on this value of 2.2 kpc. The OH spectra show absorption lines
at velocities less than +7.6 \k \  and emission lines above this velocity.
If the systemic velocity of Puppis A is indeed +7.6 \k, then the kinematic
distance would be 1.3 kpc, after applying the Galactic rotation
model of \citet{b11}.
 
To clarify this issue, we analyzed the \HI \ column densities obtained by
integrating the brightness temperature in two velocity ranges: from $v=-10$
to $+7.6$ \k , and from $-10$ to $+16$ \k. The two column densities
obtained are $N_{\rm H}=1.1$ and $2.5\times 10^{21}$ \cm2 \, respectively. 
There are no previous direct measurements of this quantity. We recall that 
these values are lower limits, since they rely on the assumption that the 
\HI \ emission is optically thin.  \citet{b1} 
analyzed the diffuse X-ray emission from Puppis A and concluded that 
$N_{\rm H}\sim 2.9\times 10^{21}$ \cm2. Fitting atmosphere models to RX 
J0822--4300, \citet{b42} obtain a 90\% confidence range of $N_{\rm H}=2.9 
- 4.7 \times 10^{21}$ \cm2, consistent with the $2 - 6\times 10^{21}$ \cm2 
\ range inferred by \citet{b37} based on X-ray emission lines of \OVII \ 
and \OVIII. From the fitted \HI \ column density, \citet{b42} infer a 
distance of between 1.9 and 2.5 kpc to Puppis A. 

The agreement of the 
higher $N_{\rm H}$ obtained by us with these previous estimates provides
supporting evidence for the choice of +16 \k \ as the systemic velocity for 
Puppis A. The only support for the lower value of $N_{\rm H}$ is the column 
density of $1.6 \times 10^{21}$ \cm2 \ estimated  by \citet{b7},  based on 
Ly$\alpha$ observations towards the nearby star HD 69882. Since this star 
is almost $1^\circ$ away from RX J0822--4300, this result may not be relevant. 
The column density estimated here when integrating from $-10$ to $+16$ \k \ 
not only supports the previously accepted kinematic distance of 2.2 kpc but 
also excludes the power-law fitted by \citet{b42} to the X-ray spectra of RX 
J0822--4300, which requires that $N_{\rm H} \sim 10^{22}$ \cm2.
 
In addition to the agreement in column densities, another argument
in favour of the systemic velocity of +16 \k \ for Puppis A is
provided by the velocity at which the double lobed \HI \ structure 
appears. \citet{b40} found that their results are compatible with 
a similar excitation temperature for the two spectral components 
in OH towards Puppis A at +3.5 and +13 \k . Given these 
assumptions, the second component must necessarily lie behind 
Puppis A in order to appear in emission. However, there is no
direct evidence to suggest that the two components do, in fact,
have the same excitation temperatures. Moreover, since the
assumed excitation temperature ($\sim 6.5$ K) is quite close to 
the brightness temperature of the continuum background (3.4 K
off source; $\sim 9$ K on source), a small change in the
excitation temperature (to $\gtrsim 10$ K), would lead to an 
emission line at +13 \k , even if the OH line arises in molecular
gas in front of the SNR. Such a condition would have been revealed 
by subtracting an ``expected'' profile from the OH spectra towards 
Puppis A.

\section {Conclusions}

We have found a second example of a CCO which appears to have 
modified the surrounding neutral gas. The CCO is RX J0822--4300
and the host SNR is Puppis A. The first example reported is 1E 
1207.4--5209 in the SNR G296.5+10.0 \citep{b21}. For both cases, 
the CCO lies near the centre of an \HI \ depression. In the case
of Puppis A, the depression consists of two lobes, located on either 
side of the CCO, with possibly a blue-shifted hole, similar to that 
found at the position of 1E 1207.4--5209. This structure appears to
be formed by the ejection of two opposing, collimated jets. This 
hypothesis is based on the assumption that the explosion site of 
Puppis A is coincident with the expansion centre of optical features 
suggested by \citet{b39}. The proper motion of RX J0822--4300 is 
estimated to be $\sim 0.1$ arcsec yr$^{-1}$. X-ray observatories 
currently in progress should be able to measure within the next 
few years if the CCO has moved from the present position, providing 
evidence to test this hypothesis. Observations of \HI \ carried out 
with higher spatial resolution, and more sensitive X-ray observations, 
would be very valuable in searches for collimated outflows ejected 
from the CCO. The present \HI \ observations have been used to 
confirm the distance of 2.2 kpc to Puppis A, derived in an earlier VLA 
\HI \ absorption study \citep{b30}. This estimate is based on a systemic 
velocity of $\sim +16$ \k \ of the double lobed structure, and reinforced  
by the agreement between the value for the \HI \ column density obtained 
by integrating the \HI \ emission from --10 to +16 \k, with those derived 
from models for X-ray spectra \citep{b37, b42}.

\section*{Acknowledgments}

We are grateful to Naomi McClure-Griffiths, who provided data from the 
SGPS and assisted with their amalgamation into the ATCA data; and to Jim 
Caswell, Pablo Vel\'azquez and Beate Woermann for helpful discussions. 
This research was partially funded through CONICET grant 4203/96, UBACYT 
grant A013 and by the Australian Research Council. The Australia Telescope 
Compact Array is funded by the Commonwealth of Australia for operation as 
a National Facility by CSIRO.
The National Radio Astronomy Observatory is a facility of the National 
Science Foundation operated under a cooperative agreement by Associated 
Universities, Inc.

\label{lastpage}

\end{document}